\documentstyle[aps,prl,epsf,floats,twocolumn,physica]{revtex}

\begin{document}

\def\gap{\scriptsize${\stackrel{\textstyle _>}{_\sim}} $\normalsize \  }

\twocolumn[\hsize\textwidth\columnwidth\hsize\csname @twocolumnfalse\endcsname

\vspace{1.3in}
\begin{center}
\today \\

\vspace{0.5 cm}

{\large {\bf Conductance Fluctuations Near the Two-Dimensional Metal-Insulator
Transition}} \\
\bigskip
K.P.\ Li$^{1}$, Dragana Popovi\'{c}$^{2}$, and S. Washburn$^{1}$ \\
\vspace{0.2 cm}
$^{1}${\it Dept.\ of Physics and
Astronomy, The University of North Carolina at Chapel Hill, Chapel Hill, NC
27599-3255, USA} \\
$^{2}${\it National High Magnetic Field Laboratory,
Florida State University, Tallahassee, FL 32306, USA} \\
%

\end{center}
\hrule
\vspace{0.2 cm}
\centerline{\bf Abstract}
\vspace{0.3 cm}

Measurements of conductance $G$ on short, wide, high-mobility Si-MOSFETs
reveal both a two-dimensional metal-insulator transition (MIT) at moderate
temperatures (1 $<~ T <$ 4~K) and mesoscopic fluctuations of the conductance
at low temperatures ($T~ <$ 1~K).  Both were studied as a function of
chemical potential (carrier concentration $n_s$) controlled by gate voltage
($V_g$) and magnetic field $B$ near the MIT.  Fourier analysis of the low
temperature fluctuations reveals several fluctuation scales in $V_g$ that vary
non-monotonically near the MIT.  At higher temperatures, $G(V_g,B)$ is
similar to large FETs and exhibits a MIT.  All of
the observations support the suggestion that the MIT is driven by Coulomb
interactions among the carriers.
\vspace{0.3 cm}

\vspace{0.5 cm}
\hrule
\vspace{1.0 cm}

]


\noindent
{\bf 1. Introduction and method}
\vspace{0.3 cm}

In spite of wide-spread acceptance of the view that two-dimensional
electron systems cannot undergo metal-insulator transitions
(MIT)~\cite{gang,weakloc}, there are now solid experimental and
theoretical reasons to accept the opposite
viewpoint~\cite{Krav,DP_MIT,Dobro}. In high-mobility ($\mu >
1$m$^2/$V$\cdot$sec) silicon metal-oxide-semiconductor field-effect
transistors (MOSFETs)~\cite{AFS}, where the disorder (fluctuations in
the impurity potential energy $W$) is small enough, a MIT is observed.
At zero magnetic field $B$, the conductivity fits the scaling
equations appropriate for quantum phase
transitions~\cite{QPT_RMP} as a function of carrier concentration and
as a function of voltage bias~\cite{Krav,DP_MIT}.
For a system in which the
Coulomb energy $U=e^2/r$ dominates (we estimate here that $U
\gg E_F $ \gap$ W$), we expect the screening length to be particularly
important and to exhibit anomalies near the MIT.
$U$ should increase as the MIT is approached from the insulating side
as separation of the electrons decreases, and from the metallic side
as the screening becomes less
efficient~\cite{AFS}.
Therefore, $U$ should be {\em maximum} near the MIT.

In very short FETs, where $L$ is comparable to relevant length scales
in the quantum transport problem, we expect to see anomalies
associated with the competing length scales~\cite{LR,QPT_RMP}.
In addition, there are complications associated with the reduced available
phase space in the small samples.  As $T \rightarrow 0$, we expect $G$
to be dominated by resonance tunneling through a few localized
states~\cite{Kirp} or by hopping along chains of such
sites~\cite{glazman}. Studies of the typical conductance $\langle\ln
G\rangle$ at low temperatures have already provided counter-intuitive
results~\cite{Pop}. At higher temperatures, we expect thermal
averaging to remove (average away) many of the fluctuations and leave
mainly the effects of bulk conductivity.  Here the generic signatures
(scaling of $\sigma(T,V_g)$, $(\sigma(V_g,B)-\sigma(V_g,0)> 0$) of the
MIT appear in large samples and will appear in ours.

Our samples were short ($L < 4~\mu$m), wide ($w > 11~\mu$m)
n-channel MOSFETs on the (100) surface of Si ($\approx 3\times
10^{20}$~acceptors/m$^{3}$, 50~nm gate oxide, oxide charge $<
10^{14}$~m$^{-2}$).  High peak mobilities ($\simeq 1$~m$^2/$V$\cdot$sec)
indicate that there is much less disorder than in samples used to
``confirm''~\cite{weakloc} the non-interacting scaling picture.  The
conductance $G(V_g,B,T)$ was measured by standard lock-in techniques
in a dilution refrigerator in a shielded enclosure at low source-drain
bias ($V_{SD} < 2 \mu V$).  For each measurement we infer a
conductivity $\sigma=(L/w)\exp <\ln G>$, and we find several anomalies 
in the region of conductivities where the MIT occurs.

\vspace{0.8 cm}
\noindent
{\bf 2. Metal-insulator transition}
\vspace{0.3 cm}

\begin{figure}[t]
\centerline{\epsfxsize=2.7in \epsfbox{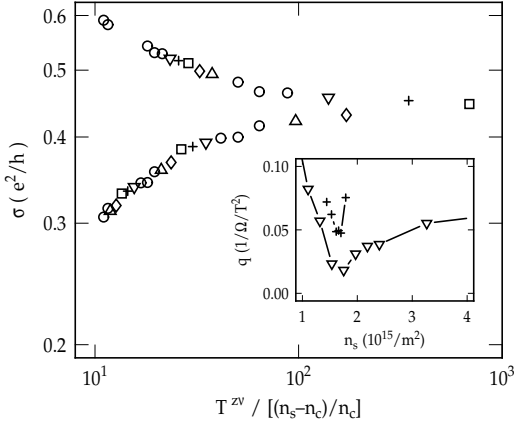}\vspace{5pt}}
\caption{Scaled conductance data for a 1.25 $\mu$m long 11.5 $\mu$m wide 
MOSFET for 1.9~K$ < T < 4.0$~K.  Each symbol represents conductance data from 
a particular temperature. 
Inset: coefficient $q$ for the parabolic $G(B)$ for the short MOSFET 
($\bigtriangledown$) at $T=0.7$~K and 
representative data from a large sample with $L=400~\mu$m ($+$) at $T=1.4$~K.
\label{scaling_plot}}
\end{figure}

The MIT appears in our samples as a crossing of the $G(V_g)$ curves at
$n_c \simeq 1.7 \times 10^{15}/$m$^2$
for a family of curves at different
temperatures.  We forgo displaying the bare $G(T)$ which are less
informative and leap immediately to the scaling plot of $\sigma$ {\it
vs.} $(T/ |\delta _{n}|)^{z \nu}$, where $\delta = (n_s - n_c) / n_c$ is
the scaled distance from the critical point and $\nu$ and $z$ are
exponents describing the scaling of the spatial and temporal
correlations, respectively, in the 2DES.  Figure~\ref{scaling_plot}
contains conductance against a scaled temperature for one of our
samples ($L= 1.25~\mu$m and $w = 11.5~\mu$m).  It is clear that the
scaling of the data provides two branch curves that contain all of the
data as expected from the theoretical arguments~\cite{Dobro,QPT_RMP}
and in accord with previous experiments~\cite{Krav,DP_MIT,SiGe_MIT} in
many systems.  We point out that $n_c=1.7\times 10^{15}/$m$^2$ agrees
well with all results on large
samples~\cite{Krav,DP_MIT,SiGe_MIT,Pudalov}, but that the exponent
$z\nu = 16 \pm 6$ is {\it much} larger than has been observed 
before.

In addition to $G(V_g,0,T)$, we have studied $G(V_g,B,T)$ for $0 < B <
2$~T.  It was learned recently that the application of $B$
perpendicular to the sample plane allows the differentiation between
terms associated with generic (non-interacting) weak localization and
with (Hartree) electron-electron collisions.  $G(B)$ can be decomposed
as $p f(B) - q B^2$, with $f(B),~ p,~ q > 0$, since the weak localization term
is expected to be positive and Hartree term is expected be negative
and quadratic up to substantial fields.  The coefficient $q$ has a
distinct minimum near the MIT~\cite{DP_MIT}. (A parallel $B$ has a
much more dramatic effect in the metallic region~\cite{Krav,Pudalov}.)
The coefficient of the Hartree terms for 
one of our samples are
plotted in the inset of Fig.~\ref{scaling_plot} along with data from a
much larger MOSFET~\cite{DP_MIT}.  Another MOSFET with $L=1.5~\mu$m 
yielded very similar results and $q$ for both samples were not 
temperature dependent down to $T=0.04$~K.   
There is a clear (but softer) minimum near the MIT in the short MOSFETs.

For the short MOSFET, 
the curve for $q$ is much flatter -- at least in the metallic region, 
which leads us to speculate that
the correlation range from the Coulomb interactions that are
generating the metallic behavior are being cut off by the sample
length and therefore ``quenching'' the more dramatic metallic behavior
seen in the large samples.  This suggestion of a cut-off is supported
by our observations that the low temperature ($T < 1.5$~K) curves of
$G(V_g)$ are largely temperature independent and {\it do not} obey the
scaling law above and in Fig.~\ref{scaling_plot}.  From analogy with
non-interacting weak localization ideas, we expect the coherence
lengths $L_\varphi$ for the quantum correlations to grow as the
temperature decreases because thermal and inelastic processes are
weaker~\cite{QPT_RMP}, so that eventually $L<L_\varphi$ and the MIT is
``short-circuited''.

\vspace{0.8 cm}
\noindent
{\bf 3. Conductance fluctuations}
\vspace{0.3 cm}

\begin{figure}[t]
\centerline{\epsfxsize=2.7in \epsfbox{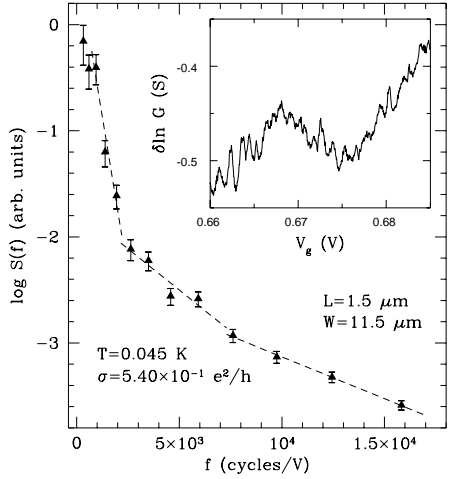}\vspace{12pt}}
\caption{Representative
power spectrum $S(f)$ of fluctuations $\delta\ln G$ for $L=1.5~\mu$m;
$W=11.5~\mu$m; $T=0.045$~K.  Dashed lines illustrate the three different
decay rates of the spectrum (from the left) $v_l$, $v_i$ and $v_h$.
The raw data for $G(V_{g})$ appear in the inset. $\sigma ~=~ 5.40\times 
10^{-1}e^{2}/h$.
\label{specs2}}
\end{figure}

As illustrated in Fig.~\ref{specs2}, $G$ can fluctuate on up to three
different scales in $V_g$~\cite{Pop,DP_RC}. The different scales are
apparent in the power spectrum $S(1/\Delta V_{g})$ [related by the
Wiener-Khinchin equations to the autocorrelation function $C(\Delta V_g)$] of
the fluctuations $\delta\ln G=\ln G(V_{g})-\langle\ln G(V_{g})\rangle$ as
three separate slopes in the low, intermediate and high ``frequency''
regions.
More than one thousand such traces have been recorded and analyzed for
this experiment.  As discussed below, certain features of the
conductance fluctuations behave regularly near the MIT and some
exhibit anomalies that might presage critical behavior in some of the
transport variables, as energy and length scales fluctuate in ways that
are completely unexpected from the point of view associated
with non-interacting electrons.

\vspace{0.8 cm}
\noindent
{\bf 4. Fluctuation correlation scales}
\vspace{0.3 cm}

The different regions of exponential decay, $S(f)=S(0)\exp (-2\pi
vf)$, correspond to contributions from distinct Lorentzians $C(\Delta
V_{g})$ of widths $\sim v_x$ (with $x$=$i$, $h$ or $l$).  The voltage
scales $v_{x}$ are related to the typical spacing of the fluctuations
for the particular transport process that leads to that Lorentzian,
and so to typical energies or lengths in the transport problem.  The
1/$v_x$ are separate fits to the different parts of $S(f)$.  The
different regions of the spectrum are separated by other 
characteristic voltage
scales.  The voltage scale for cross-over between $v_l$ and the rest
of the spectrum, does not depend on $V_g$ or $G$ $v_{0}=0.45\pm
0.15$~mV.  Similarly $v_h \simeq 50~\mu$V independent of measurement
parameters.  The remaining correlation scales, however, vary {\it
non-monotonically}
by substantial amounts near the MIT.  For example, $v_i$ emerges only in
short samples at low temperatures near the MIT.  {\em Else it vanishes.}

\begin{figure}[t]
\centerline{\epsfxsize=2.7in \epsfbox{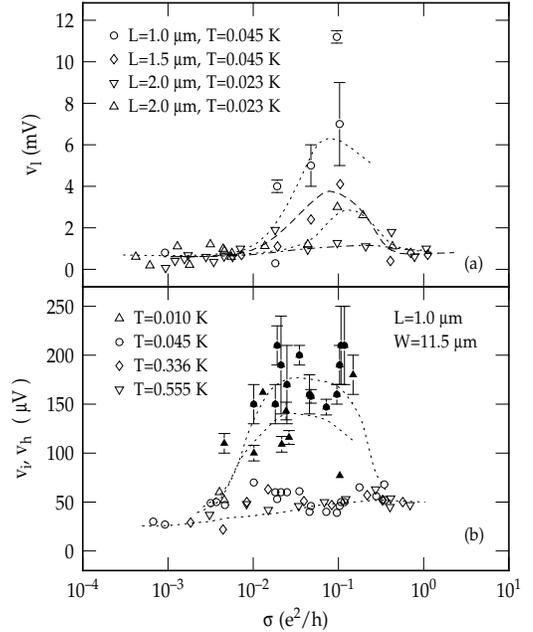}\vspace{6pt}}
\caption{(a) 
Correlation voltage $v_{l}$ for several samples at low temperatures.  
The various lines guide the eye.  One sample ($\bigtriangledown$) was 
162 $\mu m$ wide, and the rest were 11.5 $\mu m$ wide.
(b) 
Correlation voltages $v_h$ (open symbols)
and $v_i$ (solid symbols)
{\it vs. }$\sigma$ for several MOSFETs.
Dashed lines guide the eye.
\label{vh}}
\end{figure}

Figure~\ref{vh}(b) contains $v_{h}$ (lower curves) and $v_{i}$(upper curves) 
for several MOSFETs.
The highest frequency scale $v_h$ is essentially constant throughout
the transition region.  This is typical of the behavior expected {\it
for all energy and length scales} in the non-interacting pictures
proposed in the original scaling model~\cite{gang}, where all
quantities should behave regularly: "smooth" and monotonic in energy.
Instead of remaining ``flat'' or regular near the MIT, $v_i$ rises by
400\% and then falls back to the original value.  A similar anomaly on
a completely different absolute energy (or length) scale (and
presumably from a different transport process) appears in $v_l$ 
[Fig.~\ref{vh}(a)].

To relate the measured fluctuation scales to energy or length scales
in the sample physics is simple in the naive sense.  A non-interacting
model of the device electrostatics provides that~\cite{AFS}
$d\mu/dV_{g}= (d\mu/dn_{s})(dn_{s}/dV_{g})=(1/D(E))(C/e)$ with $C=
6.92\times 10^{-4}$~F/m$^{2}$ as the gate capacitance per area and
$D(E)$ as the density of states.  It is mainly owing to lack of a
reliable model for $D(E)$ that 
serious correspondence between the electrostatics (which we expect to be
largely reliable even near the MIT) and the electron energies in the
MOSFETs can not be provided.
Non-interacting pictures lead to $D(E) \simeq 1.6\times
10^{18}$~m$^{-2}$eV$^{-1}$ for high $V_g$ and considerably reduced as
$\sigma$ falls below the MIT.  This places $v_h$ in the 1~$\mu$eV
range, which is about the same as the level spacing for the electrons,
if the effective sample area is a few $\mu$m$^2$, which is quite reasonable.
The lack of critical behavior is also in line with this association.

The other energy scales are correspondingly larger.  Without
substantial theoretical insight for this problem, we are not able to
assign them reliably to particular energy scales or to any specific
physics.  One plausible sugestion is that $v_i$, which grows from and
retreats into $v_h$ near the MIT, is a disturbance in the energy level
spacing for some of the states, perhaps as separate chains of localized
states begin to exchange carriers as the screening length passes
through a maximum.

\vspace{0.8 cm}
\noindent
{\bf 5. Magnetoconductance fluctuations}
\vspace{0.3 cm}

\begin{figure}[t]
\vspace{-40pt}
\centerline{\epsfxsize=3in \epsfbox{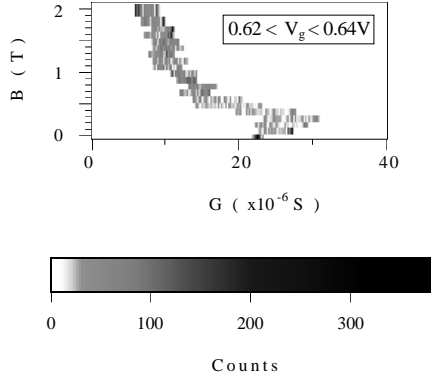}\vspace{-66pt}}
\caption{Distribution function of $G(V_g,B)$ 
for a sample with $L=3.5~\mu$m and $w=11.5~\mu$m (so 
$\sigma (0)= 0.17 e^2/h$) at $T=0.04~$K.
Conductance fluctuations appear ``on top of'' the positive and negative
components of $G(B)$ discussed above in the context of MIT at higher
temperatures.
\label{KPL_GB}}
\end{figure}

The fluctuations in $G(B)$ have been studied in depth.  Generally
speaking we find that the fluctuations are superimposed on the
high-temperature $G(B)$ discussed above.  An example of the fluctuations
appears in a topographical display of $G(V_g,B)$ in Fig.~\ref{KPL_GB}.
Signatures of both positive and negative ($p$ and $q$) components of
$G(B)$ are apparent in spite of the presence of sizable fluctuations.

\vspace{0.8 cm}
\noindent
{\bf 6. Conclusion}
\vspace{0.3 cm}

We have studied the 2d MIT in short Si-MOSFETs, where the sample
length $L$ competes with the length scales controlling the MIT.  We
find several anomalies in fluctuation correlation scales at low
temperatures $T<1$~K and a ``quenched'' MIT at higher temperatures in
which development of the metallic phase has been short-circuited as
$L$ limits the range of the important quantum correlations.  

\vspace{0.8cm}
\noindent {\bf Acknowledgements}
\vspace{0.3cm}

The authors are grateful to B.L.\ Altshuler, V.\ Dobrosavljevi\'{c}, A.B.\
Fowler, and B.B.\ Mandelbrot for useful discussions.  This work was supported
by NSF Grant DMR-9510355.

\end{document}